\documentclass[prd,aps,showpacs,nofootinbib,tightenlines]{revtex4}  %%preprint,
\usepackage{mathrsfs}
\usepackage{amsmath}
\usepackage{amssymb}
\usepackage{epsfig}
\usepackage{graphicx}
\usepackage{booktabs}
\usepackage{multirow}
\usepackage{subfigure}
\usepackage{color}
\begin{document}
%%%%%%%%%%%%%%%%%%%%%%%%%%%%%%%%%%%%%%%%%%%%%
%\renewcommand{\arraystretch}{0.5}
\newcommand{\psl}{ p \hspace{-1.8truemm}/ }
\newcommand{\nsl}{ n \hspace{-2.2truemm}/ }
\newcommand{\vsl}{ v \hspace{-2.2truemm}/ }
\newcommand{\epsl}{\epsilon \hspace{-1.8truemm}/\,  }

\title{Isovector scalar $a_0(980)$ and $a_0(1450)$ resonances in the  $B\rightarrow \psi (K\bar{K},\pi\eta) $ decays }
\author{Zhou Rui }\email{jindui1127@126.com}
\author{Ya-Qian Li }
\author{Jie Zhang }
%\author{Wen-Fei Wang$^2$}\email{wfwang@sxu.edu.cn}
\affiliation{ College of Sciences, North China University of Science and Technology, Tangshan 063009,  China}
%\affiliation{$^2$Institute of Theoretical Physics, Shanxi University, Taiyuan, Shanxi 030006, China}
\date{\today}

\begin{abstract}
We present an analysis of   two isovector scalar resonant contributions to the
 $B$  decays into charmonia plus $K\bar K$ or $\pi\eta$ pair in the perturbative QCD approach.
The Flatt\'{e} model for the $a_0(980)$ resonance and  the Breit Wigner formula for the $a_0(1450)$ resonance
 are adopted to parametrize the timelike  form factors in the dimeson distribution
amplitudes, which capture the important final state interactions in these processes.
 The predicted distribution in the $K^+K^-$ invariant mass  as well as
 its integrated branching ratio for the $a_0(980)$ resonance in the
 $B^0\rightarrow J/\psi K^+K^-$ mode agree well with the current available experimental data.
The obtained branching ratio of the quasi-two-body decay $B^0\rightarrow J/\psi a_0(980)(\rightarrow \pi^0\eta)$
  can reach the order of $10^{-6}$,  letting the corresponding measurement appear feasible.
For the  $a_0(1450)$ component,
our results could be tested by further experiments in the LHCb and Belle II.
 We also discuss some theoretical uncertainties in detail in our calculation.
\end{abstract}


%\keywords{ $B_c$ meson, Weak decays, $P$-wave charmonium, Branching ratio}

\maketitle

\section{Introduction}
Many scalar mesons with quantum numbers $J^P=0^+$  have been well established in the experiment \cite{pdg2016}.
 Amongst them, two important low-lying  scalar resonances, namely, isoscalar $f_0(980)$ and isovector $a_0(980)$,
  are of special interest. Their almost degenerate masses
would lead to a mixing with each other through isospin violating effects \cite{prd75114012,prd76074028,prd92034010,plb759501}.
As  their masses proximity to  the $K\bar{K}$ threshold, both   can strong coupling to $K\bar{K}$.
 Besides, their main individual decay chain are  $f_0(980)\rightarrow \pi\pi$ and $a_0(980)\rightarrow \pi\eta$, respectively.
Up to now, several $B$ decays involving scalar mesons have been observed, either with an
$f_0(980)$ \cite{prd72072003,prl96251803} or $a_0(980)$ \cite{prd70111102} in the final state.
Most recently, the BESIII Collaboration  reports the first observation of $a_0(980)$
meson in the semileptonic decay $D^0\rightarrow a_0(980)^-e^+\nu_e$ \cite{prl121081802},
 which provides one more  arena
in the investigation of the nature of the puzzling $a_0(980)$ states.
Above the $a_0(980)$ mass, another important isovector scalar state, $a_0(1450)$,
had been observed in $p\bar p$ annihilation experiments \cite{cbc1,cbc2}
 and  the three-body $D$ decays \cite{prd78072003,prd93052018}.
 Measurements of $B$ decays into a scalar meson can provide valuable
 information on constraining any phenomenological models
trying to understand the nature of scalar mesons.
In the quark model scenario,
the composition of $f_0(980)$ and $a_0(980)$ have turned out to be mysterious.
Their intriguing internal structure allows tests of various hypotheses, such as
quark-antiquark \cite{npb315465}, tetraquarks \cite{prd27588},
$K\bar{K}$ molecule \cite{prd412236}  and hybrid states \cite{uk1995}.
 In contrast to the unclear assignment of $a_0(980)$, it is widely accepted that $a_0(1450)$ is
  the isovector scalar $q\bar{q}$ ground state \cite{pdg2016}. In particular,
  the lattice QCD calculations support that the lowest isovector
scalar $q\bar{q}$ state corresponds to $a_0(1450)$ rather than
 $a_0(980)$ \cite{prd70094503,prd73094505,prd76114505}. For recent
 lattice QCD studies of light scalar mesons,  refer to \cite{hsc1,hsc2,prd97034506}.
 %have been published in Refs.
%in the last couple of years

From the theoretical perspective,
studies of the three-body decays of the $B$ meson with final states including  a $J/\psi$
will help us to clarify the nature of the resonances involved.
In Ref. \cite{epjc75609}, the $B_{(s)}$ decay into $J/\psi$ plus $K\bar{K}$ or $\pi\eta$ pair are studied
 by the chiral unitary approach, where the $K\bar{K}$ and  $\pi\eta$ mass distributions
 are calculated for the relevant processes.
More  general review about the use
of the chiral unitary approach to study the final-state strong interactions in weak decays,
one refer to \cite{ijmpe25} for details.
 It is found both $f_0(980)$ and $a_0(980)$ resonances contribute to the $B^0\rightarrow J/\psi K^+K^-$,
 while only the $f_0(980)$ ($a_0(980)$) resonance influences the distribution in $B_s\rightarrow J/\psi K^+K^-$
 ($B^0\rightarrow J/\psi \pi^0\eta$).
 The obtained results compared reasonably well with present experimental information.
 In Ref. \cite{jhep04010}, the authors extract  information on $\pi\eta$ scattering
 through the $B^0\rightarrow J/\psi(\pi\eta,K\bar{K})$ decays by the dispersion theory.
 The dimeson scalar form factors are introduced to describe the $S$-wave decay amplitude for the considered processes.
 The predicted decay rates are  of the same order of magnitude as those of  $\pi\pi$ analogues.
Experimentally,  evidence of the  $a_0(980)$ resonance is reported with statistical
significance of 3.9 standard deviations in the $B^0\rightarrow J/\psi K^+K^-$ decay by the LHCb Collaboration \cite{prd88072005}.
The product branching fraction of the $a_0(980)$ resonance mode is measured for the first time, yielding
\begin{eqnarray}\label{eq:mllt}
\mathcal{B}(B^0\rightarrow J/\psi a_0(980)(\rightarrow K^+K^-))=(4.70\pm 3.31\pm 0.72) \times 10^{-7},
\end{eqnarray}
where the first uncertainty is statistical and the second is systematic.

The perturbative QCD (pQCD) approach \cite{prl744388,plb348597}  is one of the recently developed theoretical tools
based on QCD to deal with various  exclusive processes \cite{pqcds}.
In our previous papers \cite{epjc77199,prd97033006},
the $S$-wave $\pi\pi(K\pi)$  resonant contributions are studied in the
 $B\rightarrow J/\psi \pi\pi(K\pi)$ decays as well as the $\psi(2S)$ counterparts.
%by using the perturbative QCD (PQCD) approach
The related scalar resonance candidates include  $f_0(500)$, $f_0(980)$, $f_0(1500)$,  $f_0(1790)$, $K^*_0(1430)$, and so on.
More recently, we studied the $P$-wave resonances, such as $\rho(770)$, $\rho(1450)$, and $\rho(1700)$,
in the $\pi^+\pi^-$ channel \cite{180904754}.
In the present paper, we mainly focus on the   isovector scalar resonances $a_0(980)$ and $a_0(1450)$
 in the $B\rightarrow \psi (K\bar K,\pi\eta)$  decays with charmonia $\psi=J/\psi,\psi(2S)$,
while the corresponding $B_s$ decay modes are forbidden because the  $s\bar s$ pair that
has $I=0$ and does not allow the isovector resonance production upon hadronization.
The subjects related to %In addition,
the crossed-channel such as $\psi P$ with $P=K,\pi,\eta$  and other higher partial wave
are beyond the scope of the present analysis.
%In particular, the $P$-wave component is very small since the dominant $P$-wave $\rho(770)$ and $\phi(1020)$ resonances
%suffer large suppression. The former is below the $K\bar K$ threshold, while the latter is forbidden
% by the Okubo-Zweig-Iizuka (OZI) rule.

As is well known the QCD
dynamics for the  three-body $B$ decays are much more complicated than those of two-body ones,
and the energy release scale for the $b$ quark mass may be too low to allow for a complete
 factorization in the central region of the Dalitz plot (DP) \cite{prd88114014,prd94114014,npb899247}.
However, based on the experimental fact that the  three-body decays of $B$ and $D$ mesons
clearly receive important contributions from intermediate resonances \cite{pdg2016},
%However, we restrict ourselves to specific kinematical configurations in which
%the three mesons are quasi aligned in the rest frame of the B.
one can assume that two of the three final-state mesons form a collimated meson pair, %originating from quark-antiquark  bilinears,
 which is interpreted as an intermediate quasi-two-body final state, and
in this case the factorization can be applied.
Within the quasi-two-body approximation,  the dominant kinematic region is restricted to the edges of a Dalitz plot,
where the three daughter mesons are quasialigned in the rest frame of the parent particle. % mother particle.
%which allows to describe three-body decays as quasi-two-body ones.

Taking the decay $B\rightarrow J/\psi K\bar K$ for example, the dominant contributions
come from the kinematic region, where the two light kaon mesons move almost parallelly for producing a resonance.
The final state interactions between the bachelor particle $J/\psi$  and the kaon pair are expected to be suppressed in such conditions.
The inherently nonperturbative dynamics associated with the kaon pair can be   parametrized into   the complex timelike
form factors involved in the two-kaon distribution amplitudes (DAs).
For the $K\bar K$ form factors,
we adopt the form as a linear combination of the $a_0(980)$ and $a_0(1450)$ resonances,
where the former is described by the popular Flatt\'{e} mass shapes based on the coupled
channels $\pi\eta$ and $K\bar K$ \cite{plb63224}, while the latter refer to the relativistic Breit-Wigner (BW) form.
\footnote{Although the width of $a_0(1450)$ is somewhat large, the  parametrizations of its resonant effect is still controversial.
We further note that most  resonances including  $a_0(1450)$  are %commonly
widely described using a relativistic BW parametrization by several collaborations in the Dalitz-plot analysis
 for the three-body $B/D$  decays \cite{prd78072003,prd93052018,prd85112010}.
 Hence, here we also model the $a_0(1450)$ by a simple BW line shape with an energy dependent width.}
The complex coefficient for each resonance are extracted
from the  isobar model fit results for the $D^0\rightarrow K^0_SK^-\pi^+$
mode performed  by the LHCb experiment \cite{prd93052018}.
Following the steps of Refs. \cite{prd91094024,epjc77199,prd97033006},
the decay amplitude for the decays under investigation  can be conceptually written as %the convolution
\begin{eqnarray} \label{eq:fac}
\mathcal{A}=\Phi_B \otimes H\otimes \Phi_{K\bar K} \otimes \Phi_{\psi},
\end{eqnarray}
where $\Phi_B$ and $\Phi_{\psi}$ are the  $B$ meson  and charmonium DAs, respectively.
The two-kaon DA $\Phi_{K\bar K}$   absorbs the nonperturbative
dynamics of the hadronization processes in the $K\bar K$ system.
The hard kernel $H$, similar to the case of two-body decays,  includes the leading-order contributions
plus the vertex corrections.
The symbol $\otimes$ denotes the convolution in parton momenta
of all the perturbative and nonperturbative objects.

The paper is structured as follows. In Sec. \ref{sec:framework},
the elementary kinematics, the  distribution amplitudes of initial and final states, and the required isovector scalar form factors are described.
In Sec. \ref{sec:result}, we present a discussion following the presentation of the significant results on the branching ratios.
 Finally, Sec. \ref{sec:sum} will be the conclusion of this work.
\section{ framework}\label{sec:framework}
\begin{figure}[!htbh]
\begin{center}
\vspace{1.5cm} \centerline{\epsfxsize=7 cm \epsffile{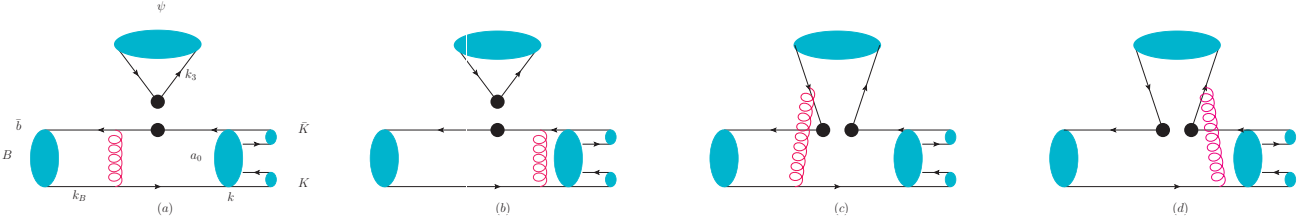}}
%\vspace{-4.5cm}
\caption{The leading-order Feynman diagrams for the
quasi-two-body decays  $B\rightarrow \psi a_0(\rightarrow K\bar K)$.
 (a,b) The factorizable  diagrams,  and (c,d) the nonfactorizable diagrams.
$a_0$ is one of the   isovector scalar intermediate states.}
 \label{fig:femy}
\end{center}
\end{figure}
We consider the decay $B\rightarrow J/\psi K\bar K$ as an illustration,
where $K\bar K$ can be either a neutral or a charged kaon pair.
In what follows, we will use the abbreviation  $a_0$ to denote the $a_0(980)$ and $a_0(1450)$ for simplicity.
It is convenient to work in the rest frame of the $B$ meson.
Its momentum $p_B$, along with the charmonium meson momentum $p_3$, the kaon pair momentum $p$ %pictorially
and other quark momenta $k_i$ in each meson, which are shown diagrammatically in  Fig. \ref{fig:femy} (a), are chosen as \cite{180904754}
\begin{eqnarray}
 p_B&=&\frac{M}{\sqrt{2}}(1,1,\textbf{0}_{T}),\quad p_3=\frac{M}{\sqrt{2}}(r^2,1-\eta,\textbf{0}_{T}),\quad  p=\frac{M}{\sqrt{2}}(1-r^2,\eta,\textbf{0}_{T}),\nonumber\\
 k_B&=&(0,\frac{M}{\sqrt{2}}x_B,\textbf{k}_{BT}),\quad k_3=(\frac{M}{\sqrt{2}}r^2x_3,\frac{M}{\sqrt{2}}(1-\eta)x_3,\textbf{k}_{3T}),\quad  k=(\frac{M}{\sqrt{2}}z(1-r^2),0,\textbf{k}_{T}),
\end{eqnarray}
with the mass ratio $r=m/M$,  and  $m(M)$ is the mass of the charmonium ($B$) meson,
the variable $\eta=\omega^2/(M^2-m^2)$,  and the invariant mass squared $\omega^2=p^2$
for the kaon pair. The individual kaon momentum $p_1$ and $p_2$ in the $K\bar K$ pair are defined as
\begin{eqnarray}
 p_1=(\zeta p^+, \eta (1-\zeta)p^+,\omega \sqrt{\zeta(1-\zeta)},0),\quad
 p_2=((1-\zeta)p^+, \eta \zeta p^+,-\omega \sqrt{\zeta(1-\zeta)},0)
\end{eqnarray}
with $\zeta$ being the kaon momentum fraction. The momenta satisfy the momentum conservation $p=p_1+p_2$.
The three-momenta of the kaon and    charmonium in the $K\bar K$ center of mass  are given by
\begin{eqnarray}
|\vec{p}_1|=\frac{\lambda^{1/2}(\omega^2,m_K^2,m_{K}^2)}{2\omega}, \quad
|\vec{p}_3|=\frac{\lambda^{1/2}(M^2,m^2,\omega^2)}{2\omega},
\end{eqnarray}
respectively, with $m_K$  the kaon  mass and  the K$\ddot{a}$ll$\acute{e}$n function $\lambda (a,b,c)= a^2+b^2+c^2-2(ab+ac+bc)$.
We do not spell out the kinematic relations for the $\pi\eta$ final state explicitly,
inasmuch as they can be obtained from the above in a straightforward manner.

In the course of the PQCD calculations, the necessary inputs contain  DAs of the initial and final states.
The $B$ meson can be treated as a heavy-light pseudoscalar system,
the structure $\gamma_{\mu}\gamma_5$ and $\gamma_5$ components remain as leading contributions.
Then, the $B$ meson wave function with an intrinsic $b$ (the conjugate space coordinate to $k_T$)
dependence  can be expressed by \cite{ppnp5185}
\begin{eqnarray}
\Phi_{B}(x,b)=\frac{i}{\sqrt{2N_c}}[(\rlap{/}{p_B}+M)\gamma_5\phi_{B}(x,b)],
\end{eqnarray}
with $N_c$ the color factor.
The DA $\phi_{B}(x,b)$ is adopted in the conventional form  \cite{ppnp5185,prd65014007}
\begin{eqnarray}
\phi_{B}(x,b)=N x^2(1-x)^2\exp[-\frac{x^2M^2}{2\omega^2_b}-\frac{\omega^2_bb^2}{2}],
\end{eqnarray}
with the shape parameter $\omega_b=0.40\pm 0.04$ GeV related to the factor $N$ by the normalization
\begin{eqnarray}
\int_0^1\phi_{B}(x,b=0)d x=\frac{f_{B}}{2\sqrt{2N_c}}.
\end{eqnarray}
Since the concerned meson pair forms a spin-0 intermediate state, the final system only has a longitudinal component.
For the  charmonium states,  the longitudinal polarized DAs are defined as \cite{prd90114030,epjc75293}
\begin{eqnarray}
\Phi_{\psi}^L=\frac{1}{\sqrt{2 N_c}}[m \rlap{/}{\epsilon_L}  \phi^L (x,b)+\rlap{/}{\epsilon_L} \rlap{/}{p_3}\phi^t (x,b)],
\end{eqnarray}
with the longitudinal polarization vector $\epsilon_L=\frac{1}{\sqrt{2}r}(-r^2,1-\eta, \textbf{0}_{T})$.
The  expressions of the  $\phi^{L,t}$  are not shown here for the sake of brevity and can be found in Refs. \cite{prd90114030,epjc75293}.

The isovector scalar DAs are introduced
in analogy with the case of two-pion ones \cite{prd91094024,plb561258},
 which are organized into
\begin{eqnarray}\label{eq:fuliye2}
\Phi_{K\bar K(\pi\eta)}^{I=1}=\frac{1}{\sqrt{2N_c}}[\rlap{/}{p}\phi^{I=1}_{v\mu=-}(z,\zeta,\omega^2)+
\omega\phi_s^{I=1}(z,\zeta,\omega^2)+\omega(\rlap{/}{n}\rlap{/}{v}-1)\phi^{I=1}_{t\mu=+}(z,\zeta,\omega^2)],
\end{eqnarray}
where $n=(1,0,\textbf{0}_{T})$  and $v=(0,1,\textbf{0}_{T})$ are two dimensionless vectors.
For $I=1$, $\phi^{I=1}_{v\mu=-}$ contributes at twist-2, while $\phi_s^{I=1}$ and $\phi^{I=1}_{t\mu=+}$
contribute at twist-3.
It is worthwhile to mention that the concerned isovector scalar dimeson systems have similar asymptotic
 DAs as the ones for a light scalar
meson \cite{prd97033006,plb730336}, but we replace the scalar decay constants with the timelike form factor:
\begin{eqnarray}\label{eq:phi0st}
\phi^{I=1}_{v\mu=-}(z,\zeta,\omega^2)&=&\phi^0=\frac{9}{\sqrt{2N_c}}F_s(\omega^2)B_1z(1-z)(1-2z), \nonumber\\
\phi_s^{I=1}(z,\zeta,\omega^2)&=&\phi^s=\frac{1}{2\sqrt{2N_c}}F_s(\omega^2),\nonumber\\
 \phi^{I=1}_{t\mu=+}(z,\zeta,\omega^2)&=&\phi^t=\frac{1}{2\sqrt{2N_c}}F_s(\omega^2)(1-2z),
\end{eqnarray}
which are the same as the two-pion one in \cite{prd91094024}, except for
the different Gegenbauer moment $B_1$ due to the SU(3) breaking effects.
%Here we assume that they are the same
Here we use $B_1=0.3$ for both $K\bar K$ and $\pi\eta$ pairs in the numerical analysis,
which is determined from the data for the $B^0\rightarrow J/\psi a_0(980)(\rightarrow K^+K^-)$ branching ratio \cite{prd88072005}.

As mentioned in the Introduction, the isovector scalar  form factors for the concerned dimeson  systems
are given by the coherence summation of the two resonances $a_0(980)$ and $a_0(1450)$,
\begin{eqnarray}\label{eq:reso}
F^{I=1}_{K\bar K(\pi\eta)}(\omega^2)=\sum_{a_0}  C_{a_0} M_{a_0}(\omega^2),
%F^{I=1}_{K\bar K}(\omega^2)= C_{KK} M_{a_0(980)}(\omega^2)+C_{KK} M_{a_0(1450)}(\omega^2),
\end{eqnarray}
where $C_{a_0}=|C_{a_0}| e^{i\phi_{a_0}}$ is the corresponding complex amplitude for each intermediate state $a_0$.
%and is the same  to the charged and neutral resonant due to the isospin symmetry.
For the $K\bar K$ pair,
the magnitude $|C_{a_0}|$ and phase $\phi_{a_0}$ can be obtained through a fit
to the data, as done successfully in Ref. \cite{prd93052018}.
As mentioned in Ref.  \cite{prd93052018},
the relevant parameters can be    fixed in the isobar model fits
in both the $D^0\rightarrow K^0_SK^-\pi^+$ and $D^0\rightarrow K^0_SK^+\pi^-$ modes,
 and two amplitude models have been constructed for each decay mode.
Because the former has a higher signal yields and smaller mistag rate with respect to the latter,
 we prefer to the experimental solution
\begin{eqnarray}\label{eq:caa}
|C_{a_0(980)}|=1.07, \quad \phi_{a_0(980)}=82^{\circ}, \quad
|C_{a_0(1450)}|=0.43, \quad \phi_{a_0(1450)}=-49^{\circ},
\end{eqnarray}
which are taken from Table V  of Ref. \cite{prd93052018}.
%It is worthwhile to stress that the above parameters are
%unsuitable for the $\pi\eta$ case because of the significant SU(3) breaking effect.
Since the experimental information on the $\pi\eta$ pair is not yet available,
in this study, we roughly estimate their magnitudes $|C_{a_0}|$ by comparing the two
 form factors of the $K\bar K$ and $\pi\eta$ pairs.
To achieve this,
taking  account of the charged dimeson pairs  $\bar {K}^0K^+$ and $\eta\pi^+$, the relevant form factors
$F_{K\bar K(\pi\eta)}(s)$,
which enter the  matrix elements for the transition from vacuum to the
 corresponding meson pairs via a $\bar u d$ source, are defined as \cite{prd96113003,epjc75488} %,prd94034008,jhep12027}
\begin{eqnarray}\label{eq:kk}
\langle \bar{K}^0K^+|\bar{u}d|0\rangle&=&B^0F_{K\bar K}(\omega^2), \nonumber\\
\langle \eta\pi^+|\bar {u} d|0\rangle&=&B^0F_{\pi\eta }(\omega^2),
\end{eqnarray}
with the scale dependence %inherent in the
factor $B^0$. %$=\frac{m_{\pi}^2}{m_u+m_d}$.
Following the  prescription in Refs. \cite{prd80054007,180902943},
%by inserting the scalar intermediate resonances propagate into above matrix elements,
by inserting a complete set of $a_0$ intermediate state into above matrix elements,
%between the currents in equat
we have
\begin{eqnarray}\label{eq:kk1}
\langle \bar{K}^0K^+(\eta\pi^+)|\bar{u}d|0\rangle_{{a_0}}
\approx\langle \bar{K}^0K^+(\eta\pi^+)|a_0\rangle \frac{1}{BW_{a_0}} \langle a_0 |\bar{u}d|0\rangle
=\frac{ g_{a_0KK}(g_{a_0\pi\eta}) \bar{f}_{a_0}m_0}{BW_{a_0}},
\end{eqnarray}
with $BW_{a_0}$ the resonance propagator \cite{160503889}. Hereafter,  $m_0$ refers to the pole mass of the resonance.
The scalar decay constant and the strong coupling constant are defined by \cite{prd88114014,prd73014017}
 \begin{eqnarray}\label{eq:coupling}
\langle a_0 |\bar{u}d|0\rangle=\bar{f}_{a_0} m_0, \quad \langle \bar{K}^0K^+(\eta\pi^+)|a_0\rangle =g_{a_0KK}(g_{a_0\pi\eta}).
\end{eqnarray}
By equating Eqs. (\ref{eq:kk}) and (\ref{eq:kk1}), we link $F_{K\bar K(\pi\eta)}(\omega^2)$  with
the usual Breit-Wigner expression through
 \begin{eqnarray}
F_{K\bar K(\pi\eta)}(\omega^2)=C^{K\bar K(\pi\eta)}_{a_0}\frac{m_0^2}{BW_{a_0}},
\end{eqnarray}
with
 \begin{eqnarray}\label{eq:ckk}
C^{K\bar K(\pi\eta)}_{a_0}=\frac{g_{a_0KK}(g_{a_0\pi\eta}) \bar{f}_{a_0}}{B^0 m_0}.
\end{eqnarray}
The combinations of $C^{K\bar K}_{a_0}$  and $C^{\pi\eta}_{a_0}$ lead to the ratio
\begin{eqnarray}\label{eq:ccc}
\frac{C^{\pi\eta}_{a_0}}{C^{K\bar K}_{a_0}} =\frac{g_{a_0\pi\eta}}{g_{a_0KK}},
\end{eqnarray}
where the values of the relative coupling $\frac{g_{a_0\pi\eta}}{g_{a_0KK}}$
for $a_0(980)$ and $a_0(1450)$ are taken from the Crystal Barrel experiment \cite{prd573860}.
It can be seen the coefficients $C_{a_0}$ have reflected the strength of the resonances $a_0$ decaying to the corresponding dimeson pair.
We then can estimate the modules of the $C_{a_0}$ for the  $\pi\eta$ system  by using  Eq. (\ref{eq:ccc}), but keep their
phases the same as in Eq. (\ref{eq:caa}).

The partial amplitude $M_R$ \footnote{ Here, we omit the relevant Blatt-Weisskopf centrifugal barrier factors
and the angular distribution factors since in the  scalar resonance
case their values  are equal to 1  \cite{prd93052018,blatt,0410014}. } appearing in Eq. (\ref{eq:reso})
are chosen depends on the resonances in question.
The $a_0(980)$  is a well established resonance but its shape is
not well described by a simple Breit-Wigner formula because
of the vicinity of the $K\bar K $ threshold.
We follow the widely accepted
prescription proposed by  Flatt\'{e} \cite{plb63224},
based on the coupled channels $\pi\eta$ and $K\bar K$.
The Flatt\'{e} mass shapes are parametrized as
\begin{eqnarray}\label{eq:flatte}
M_{a_0(980)}(\omega^2)=\frac{m_0^2}{m_0^2-\omega^2-i(g_{\pi\eta}^2\rho_{\pi\eta}+g_{KK}^2\rho_{KK})},
\end{eqnarray}
with the  nominal $a_0(980)$   mass $m_0=0.925$ GeV \cite{prd93052018}.
Note that the coupling constants $g_{KK}(g_{\pi\eta})$ in Eq. (\ref{eq:flatte}) are related
to  those in Eq. (\ref{eq:coupling}) through the relation  $g_{KK}(g_{\pi\eta})=g_{a_0KK}(g_{a_0\pi\eta})/(4\sqrt{\pi})$ according to the different definitions
between Ref. \cite{prd88114014} and Ref. \cite{prd573860}.
 %The  values of the coupling constants fitted by several Collaborations have been summarized in Ref. \cite{prd92034010}.
% Note that the coupling constants differs from
In this study, we employ parameters
$g_{\pi\eta}=0.324$ GeV and $g^2_{K\bar K}/g^2_{\pi\eta}=1.03$
from the Crystal Barrel experiment \cite{prd573860}.
The $\rho$ factors  are given by the Lorentz-invariant phase space
\begin{eqnarray}
\rho_{\pi\eta}&=&\sqrt{[1-(\frac{m_{\eta}-m_{\pi}}{\omega})^2][1-(\frac{m_{\eta}+m_{\pi}}{\omega})^2]},\nonumber\\
\rho_{K\bar K}&=&\frac{1}{2}\sqrt{1-\frac{4m^2_{K^{\pm}}}{\omega^2}}+\frac{1}{2}\sqrt{1-\frac{4m^2_{K^{0}}}{\omega^2}}.
\end{eqnarray}
The partial amplitude $M_{a_0{1450}}(\omega^2)$ picks up the conventional  Breit-Wigner model,
\begin{eqnarray}
M_{a_0(1450)}(\omega^2)=\frac{m_0^2}{m_{0}^2-\omega^2-im_{0}\Gamma(\omega)},
\end{eqnarray}
where $\Gamma(\omega)$  is its energy dependent width that is parametrized as in the case of a scalar resonance
\begin{eqnarray}
\Gamma(\omega)=\Gamma_0 \frac{|\vec{p}_1|}{|\vec{p}_{10}|}\frac{m_0}{\omega},
\end{eqnarray}
with $m_0=1.458$ GeV and $\Gamma_0=0.282$ GeV for the $a_0(1450)$ resonance \cite{prd93052018}.
The symbol $|\vec{p}_{10}|$ is used to indicate value of
$|\vec{p}_{1}|$ at the resonance peak mass.

\section{ Numerical results}\label{sec:result}
The differential branching ratio for the considered decays is explicitly written as
\begin{eqnarray}\label{eq:dfenzhibi}
\frac{d \mathcal{B}}{d \omega}=\frac{\tau \omega|\vec{p}_1||\vec{p}_3|}{32\pi^3M^3}|\mathcal{A}|^2,
\end{eqnarray}
with $\tau$  the $B$ meson lifetime.
 The resulting decay amplitudes $\mathcal{A}$ are equivalent to previous calculations in Ref. \cite{prd91094024}
 by replacing the $S$-wave $\pi\pi$ form factor
 with the corresponding $K\bar K(\pi\eta)$ one in Eq. (\ref{eq:reso}).
 To proceed with the numerical analysis, it is useful to summarize
all of the input quantities entering the PQCD approach below:
 \begin{itemize}
\item[$\bullet$] For the  masses (in GeV) \cite{pdg2016}: $M_B=5.28$, \quad $m_{J/\psi}=3.097$,
\quad $m_{\psi(2S)}=3.686$, \quad $m_{K^{\pm}}=0.494$,  \quad $m_{K^0}=0.498$,  \quad $m_{\eta}=0.548$,
\quad $m_{\pi^+}=0.14$, \quad $m_{\pi^0}=0.135$,  \quad $m_b(\text{pole})=4.8$, \quad
$\overline{m_c}(\overline{m_c})=1.275$.
\item[$\bullet$] For the Wolfenstein parameters   \cite{pdg2016}:
$\lambda = 0.22453$,\quad $A=0.836$, \quad $\bar{\rho}=0.122$,\quad $\bar{\eta}=0.355$.
\item[$\bullet$] For the  decay constants (in GeV): $f_B=0.19$ \cite{pdg2016},
\quad $f_{J/\psi}=0.405$ \cite{prd90114030}, \quad $f_{\psi(2S)}=0.296$ \cite{epjc75293}.
\item[$\bullet$] For the lifetimes (in ps) \cite{pdg2016}: $ \tau_{B_0}=1.52,\quad \tau_{B^+}=1.638$.
\end{itemize}
The  relevant parameters in the timelike form factors  have been given  in the previous section.

By using Eq. (\ref{eq:dfenzhibi}),
integrating over the full invariant mass spectrum [$\omega_{\text{min}}<\omega<M_B-m_{\psi}$ with $\omega_{\text{min}}=2m_{K^{\pm}}(m_{\pi^0}+m_{\eta})$ for $K\bar K(\pi\eta)$ modes]
separately for the individual resonant components,
we derived the $CP$-averaged branching ratios for the neutral decay modes, which
are summarized in Table \ref{tab:br}. The corresponding numbers for the charged decay modes can be obtained
by multiplying the neutral branching ratios with a factor of  $2 \tau_{B^+}/\tau_{B^0}$ in the limit of isospin symmetry.
\begin{table}
\caption{PQCD predictions for the concerned quasi-two-body decays involving the isovector scalar resonant $a_0$.
The theoretical errors correspond to the uncertainties due to the shape parameters $\omega_b$ in the wave
function of the $B$ meson,  the Gegenbauer moment $B_1$,  the magnitude of the $C_{a_0}$, and the hard scale $t$, respectively. }
\label{tab:br}
\begin{tabular}[t]{lc}
\hline\hline
Modes & $\mathcal{B}$  \\
\hline
$B^0\rightarrow J/\psi a_0(980)(\rightarrow K^+K^-)$ & $(4.7^{+1.4+1.5+1.0+1.0}_{-0.8-1.0-0.9-0.4})\times 10^{-7}$\\
$B^0\rightarrow J/\psi a_0(980)(\rightarrow  \pi^0\eta)$ & $(6.0^{+1.6+1.9+1.3+1.1}_{-1.3-1.6-1.1-0.8})\times 10^{-6}$\\
$B^0\rightarrow J/\psi a_0(1450)(\rightarrow K^+K^-)$ & $(6.8^{+3.8+0.7+1.4+0.4}_{-2.2-0.2-1.2-0.1})\times 10^{-7}$\\
$B^0\rightarrow J/\psi a_0(1450)(\rightarrow \pi^0\eta)$ & $(1.1^{+0.5+0.1+0.2+0.0}_{-0.4-0.1-0.2-0.0})\times 10^{-6}$\\
$B^0\rightarrow \psi(2S) a_0(980)(\rightarrow K^+K^-)$ & $(7.9^{+1.4+2.5+1.6+1.4}_{-1.3-1.7-1.5-0.8})\times 10^{-8}$\\
$B^0\rightarrow \psi(2S) a_0(980)(\rightarrow \pi^0\eta)$ & $(1.5^{+0.3+0.4+0.3+0.2}_{-0.3-0.4-0.3-0.2})\times 10^{-6}$\\
$B^0\rightarrow \psi(2S) a_0(1450)(\rightarrow K^+K^-)$ & $(6.1^{+4.5+0.9+1.3+0.4}_{-2.6-0.1-1.2-0.1})\times 10^{-8}$\\
$B^0\rightarrow \psi(2S) a_0(1450)(\rightarrow \pi^0\eta)$ & $(1.2^{+0.6+0.1+0.3+0.0}_{-0.5-0.1-0.2-0.1})\times 10^{-7}$\\
\hline\hline
\end{tabular}
\end{table}
For our results,  we take into account the following theoretical  uncertainties.
The first  uncertainty is from the shape parameter  in the $B$ meson wave function, $\omega_b=0.40\pm 0.04$.
The second  error originates from the Gegenbauer moment $B_1=0.3\pm 0.1$ from the twist-2 DAs in Eq. (\ref{eq:phi0st}).
While the twist-3 DAs are taken as the asymptotic form for lack
of better results from nonperturbative methods, this may also give large uncertainties.
The third error is induced by the complex parameters $C_{a_0}$ in Eq. (\ref{eq:reso}).
In the evaluation, we vary their magnitudes  within a $10\%$ range.
The last one is caused by the variation of the hard scale from 0.75t to 1.25t, which
characterizes the energy release in decay process.
It is found that the main uncertainties of the concerned
processes come from those nonperturbative parameters associated with the DAs of the $B$ meson and dimeson pair.
Their combined uncertainties can reach $50\%$.
 The uncertainties stemming from the Flatt\'{e} parameters in Eq. (\ref{eq:flatte})
for the $a_0(980)$ channels are not included in Table \ref{tab:br},
whose effect on the branching ratio will be discussed in detail later.
The errors from the uncertainty
of the CKM matrix elements and the decay constants
of charmonia are very small and have been neglected.

 We notice that the branching ratios associated with  $a_0(1450)$ modes
 are more sensitive to the shape parameter $\omega_b$ than the Gegenbauer moment $B_1$,
whereas the situation is different for the corresponding processes of $a_0(980)$.
It can be simply understood from the different twist contributions  in the dimeson DAs.
For $a_0(980)$ channels, the twist-2 and twist-3 contributions are of the same order,
while for the $a_0(1450)$ case,
the latter are  more larger than the former.
It is easy to observe that in Eq.(\ref{eq:fuliye2}),
the twist-3 DAs always multiply by the invariant mass $\omega$,
and the larger pole mass induces  larger contributions from twist-3 DAs.
As the hard amplitude in Eq. (\ref{eq:fac})  is convoluted with initial-state and final-state hadron DAs,
the twist-3 contributions  are concentrated in the endpoint region \cite{prd65014007},
which correspond to a small hard scale for the hard amplitude,
such that the running coupling constant evaluated at that scale rise up rapidly.
Therefore, the twist-3  contributions are more sensitive to the $\omega_b$,
which characterizes the shape of $B$ meson DA.
As stated above, the twist-3 DAs give the dominant contribution to the $a_0(1450)$ channels,
thus their branching ratios depend heavily on  $\omega_b$ and
 are less sensitive to the variation of the  Gegenbauer moment $B_1$,
 which appears in the twist-2 DA.

It is well known that the $a_0(980) $ resonance mass
 always near the $K\bar K$ thresholds. As a result
the predicted  branching ratio of the $B^0\rightarrow \psi a_0(980)(\rightarrow K \bar K)$ decay
is very sensitive to the choice of the  resonance mass.
It was pointed out in Refs. \cite{jpg35075005,prd78074023} that
 when we use the Flatt\'{e} parametrization for the $a_0(980) $ resonance,
there is a strong correlation between  its mass and  coupling constants.
Therefore, here we do not take into account their individual uncertainty,
but check  the sensitivity of our results to the choice of these Flatt\'{e} parameters.
Actually, several phenomenological models \cite{npb315465,prd27588,uk1995}
and experimental measurements \cite{prd78074023,plb47953,prd84112009,plb6815,prd59012001} determined the relevant
Flatt\'{e} parameters (mass and couplings) as listed in Table \ref{tab:gm0br}.
Some coupling constants are converted into the numbers according to the definition in Ref. \cite{prd573860}.
The first four parameter sets are taken  from phenomenological models, while the remainder   come from the experimental fitting.
With each parameter set, we obtain the corresponding branching ratio shown in the last column of  Table \ref{tab:gm0br}.
One can  see that   the parameters are  quite model-dependent and suffers sizable uncertainty,
which leads to the  yielding branching ratios lie in a wide  range $(2.8\sim 18)\times 10^{-7}$.
We expect that the new and improved data would help in constraining the relevant parameters
and our theoretical understanding of the properties of the $a_0(980)$ resonances.

\begin{table}
\caption{Masses and coupling constants of the $a_0(980)$ resonance in the Flatt\'{e} parametrization
 determined from  various theoretical models and  experimental data.
 The last  column  correspond to the calculated branching ratios in the PQCD approach by using the corresponding parameters.}
\label{tab:gm0br}
\begin{tabular}[t]{lcccc}
\hline\hline
Model or experiment &$m_{a_0(980)}$ (\text{MeV}) & $g_{\pi\eta}$ (\text{MeV})
 & $g_{KK}$ (\text{MeV}) & $\mathcal{B}(B^0\rightarrow J/\psi a_0(980)(\rightarrow K^+K^-))$   \\ \hline
$q\bar q$ model \cite{npb315465}  & 983 & 287 &179 &$1.7\times 10^{-6}$    \\
$qq\bar{qq}$ model \cite{npb315465}  & 983 & 645 &757 &$2.8\times 10^{-7}$   \\
$K\bar K$ model \cite{prd27588}  & 980 & 245 &386 &$1.5\times 10^{-6}$    \\
$q\bar qg$ model \cite{uk1995}  & 980 & 355 &278 &$1.2\times 10^{-6}$     \\
 CB \cite{prd78074023}  & 987.4 & 405 &415 &$9.2\times 10^{-7}$    \\
 SND \cite{plb47953} & 995& 439&592&$6.6\times 10^{-7}$    \\
 CLEO \cite{prd84112009} & 998 & 600 &396 &$5.4\times 10^{-7}$    \\
 KLOE \cite{plb6815} \footnotemark[1] & 982.5 & 303 &397 &$1.3\times 10^{-6}$     \\
 E852 \cite{prd59012001}  & 1001 & 348 &235 &$1.8\times 10^{-6}$    \\
\hline\hline
\end{tabular}
\footnotetext[1]{ We quote the fit result for the KL model. }
\end{table}

Now we turn to estimate  the isospin breaking effect between the two physics
final states $K^+K^-$ and $K^0\bar{K}^0$ in the isovector $a_0$ channels.
Since both the charged and neutral kaons decay channel open near the $a_0(980)$ resonance mass,
the 8 MeV gap between the $K^+K^-$ and $K^0\bar K^0$ thresholds make the latter mode
 suffer  a further suppression from the phase space.
Hence,  the isospin breaking effect may be non-negligible in the $a_0(980)$ channels.
In principle, as mentioned before,
 the nearly degenerate masses between the $a_0(980)$ and $f_0(980)$ resonances
would lead to an admixture of them,
which   also cause to an important  isospin-violating effects.
 However, it is not the theme of the present work.
  Here, we roughly estimate the isospin-violating effect from the kaon mass differences,
  but assume isospin symmetry for their coupling constants.
To be more specific,  we  calculated the corresponding $K^0\bar K^0$ modes by using the same input parameters
as the $K^+K^-$ ones except for distinguishing the charged and neutral kaon masses, and  found numerically that
\begin{eqnarray}
\mathcal{B}(B^0\rightarrow J/\psi a_0(980)( \rightarrow K^0\bar{K}^0))&=&4.3\times 10^{-7},\nonumber\\
\mathcal{B}(B^0\rightarrow \psi(2S) a_0(980)( \rightarrow K^0\bar{K}^0))&=&7.1\times 10^{-8},\nonumber\\
\mathcal{B}(B^0\rightarrow J/\psi a_0(1450)( \rightarrow K^0\bar{K}^0))&=&6.7\times 10^{-7},\nonumber\\
\mathcal{B}(B^0\rightarrow \psi(2S) a_0(1450)( \rightarrow K^0\bar{K}^0))&=&5.8\times 10^{-8},
\end{eqnarray}
which are typical smaller than the corresponding numbers for the charged kaon channels in Table \ref{tab:br}.
For the $a_0(980)$ channels, the isospin breaking effect can reach roughly 10 percents
 even though the $a_0(980)-f_0(980)$ mixing is not included.
For the case of $a_0(1450)$,
the isospin breaking effect  are rather small as expected since its resonance mass is far away from the two-kaon thresholds.

Let us also compare our results to those obtained in other methods.
As stated before,  the decay under investigation have been discussed in the chiral unitary approach \cite{epjc75609}
and the dispersion theory \cite{jhep04010}.
 The authors of Ref. \cite{epjc75609}  subtract a smooth but large background
 from the differential decay to get the $a_0(980)$ contribution, and estimate the value
$\mathcal{B}(B^0\rightarrow J/\psi a_0(980)(\rightarrow \pi^0\eta))=(2.2\pm 0.2)\times10^{-6}$,
which is smaller than our numbers in Table \ref{tab:br} by a factor of 3.
Nevertheless, a phenomenological estimate for the branching ratio of $B^0\rightarrow J/\psi  \pi^0\eta$
in the mass range above threshold up to 1.1 GeV in \cite{jhep04010}
gave a  range  $(6.0\sim 6.4)\times 10^{-6}$ with the input phase $\delta_{12}=90^{\circ}$.
For the sake of comparison,
we derive a central value of $5.7\times 10^{-6}$
within the same energy region [$m_{\pi}+m_{\eta},1.1 \text{GeV}$].
This is consistent with their theoretical estimates within errors.

On the experimental side,
the decay $B^0\rightarrow J/\psi K^+K^-$  is first observed by the LHCb Collaboration.
The relevant  amplitude analysis  is performed to separate resonant and nonresonant
contributions in the $K^+K^-$ spectrum. There is $3.9\sigma$ evidence for the $a_0(980)$ resonance
with a product branching fraction of
\begin{eqnarray}
\mathcal{B}(B^0\rightarrow J/\psi a_0(980)(\rightarrow K^+K^-))=(4.70\pm 3.31\pm 0.72) \times 10^{-7},
\end{eqnarray}
and an upper limit on its branching fraction is set to be $9.0\times 10^{-7}$ at $ 90\%$ confidence level.
 The measured central value is close to our result in Table \ref{tab:br},
whereas the statistical uncertainty  is too large to make a definite conclusion.
As indicated in Table \ref{tab:gm0br},
there is a notable uncertainty due to  the different solutions of the Flatt\'{e} parameters,
and some of the  results exceed the experimental upper limit.
We suggest the experimentalists carry out a more precise measurement on this channel to constrain the relevant parameters,
which allow us to discriminate between different models and improve the approach.
On the other hand,
the experiment information on the $\pi\eta$ channels are still scarcer.
As a cross-check to the dynamical calculations, using the PQCD predictions as
given in Table \ref{tab:br} we can estimate the relative
ratios ${\mathcal R}^{\psi}_{a_0}=\frac{\mathcal {B}(B^0\rightarrow \psi a_0(\rightarrow K^+K^-))}
{\mathcal {B}(B^0\rightarrow \psi a_0(\rightarrow \pi^0\eta))}$ as below,
\begin{eqnarray}\label{eq:rat}
{\mathcal R}^{J/\psi}_{a_0(980)}=0.08^{+0.04}_{-0.00},\quad {\mathcal R}^{\psi(2S)}_{a_0(980)}=0.05^{+0.01}_{-0.00}, \quad
{\mathcal R}^{J/\psi}_{a_0(1450)}=0.62^{+0.02}_{-0.04},\quad {\mathcal R}^{\psi(2S)}_{a_0(1450)}=0.51^{+0.10}_{-0.05},
\end{eqnarray}
where all uncertainties are added in quadrature.
Under the narrow width approximation, above ratios  obeys a simple factorization relation
\begin{eqnarray}
\mathcal{R}^{\psi}_{a_0}\approx \frac{\mathcal{B}(B^0\rightarrow \psi a_0)\mathcal{B}(a_0 \rightarrow K^+K^-)}
{\mathcal{B}(B^0\rightarrow \psi a_0)\mathcal{B}(a_0 \rightarrow \pi^0\eta)}
=\frac{\Gamma(a_0\rightarrow K^+K^-)}{\Gamma(a_0\rightarrow \pi^0\eta)},
\end{eqnarray}
which allows us to test the ratios in Eq. (\ref{eq:rat}).
The average values of  the relative partial decay widths
$\Gamma(a_0\rightarrow K\bar K)/\Gamma(a_0\rightarrow \pi^0\eta)$ given by Particle Data Group (PDG) \cite{pdg2016}
 for the  resonances $a_0(980)$ and $a_0(1450)$ are  $0.183\pm 0.024$ and  $0.88\pm 0.23$, respectively.
Recalling that the isospin relation $\Gamma(a_0\rightarrow K^+ K^-)=\Gamma(a_0\rightarrow K\bar K)/2$,
our calculations are in accordance with  the data within errors.

\begin{figure}[tbp]
\begin{center}
\setlength{\abovecaptionskip}{0pt}
\centerline{
\hspace{4cm}\subfigure{\epsfxsize=13 cm \epsffile{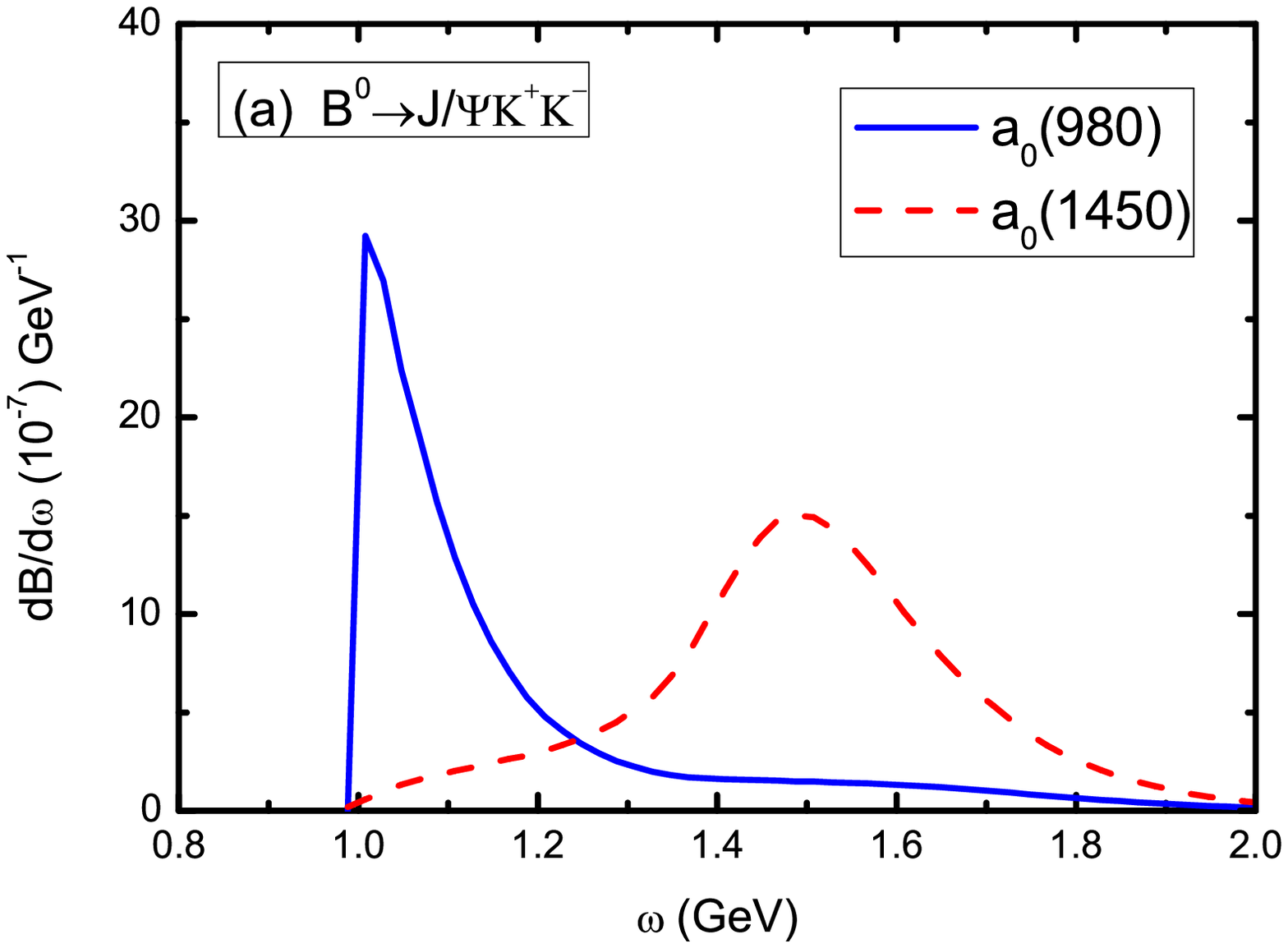} }
\hspace{-6cm}\subfigure{ \epsfxsize=13 cm \epsffile{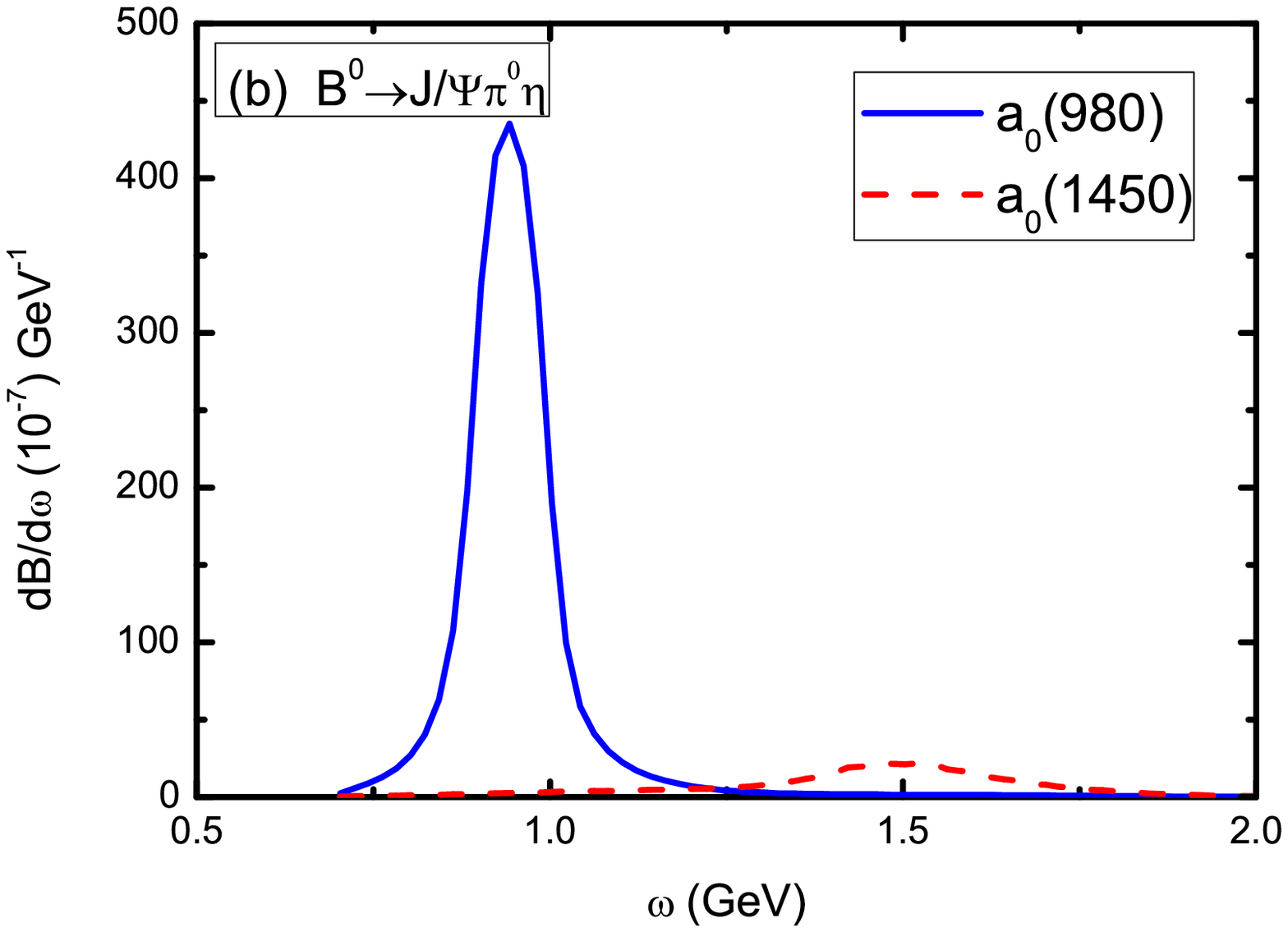}}}
\vspace{-3cm}
\centerline{
\hspace{4cm}\subfigure{\epsfxsize=13 cm \epsffile{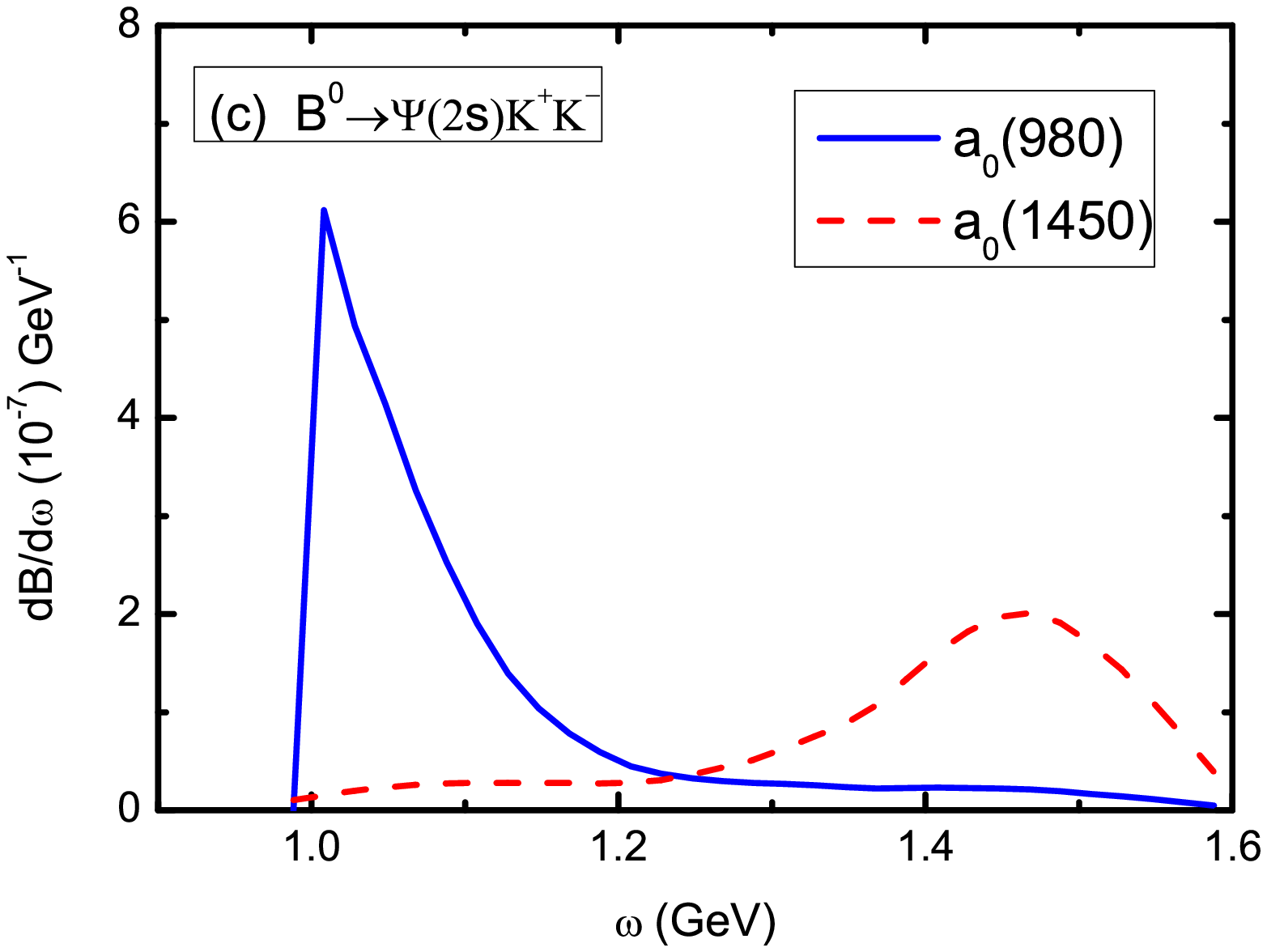} }
\hspace{-6cm}\subfigure{ \epsfxsize=13 cm \epsffile{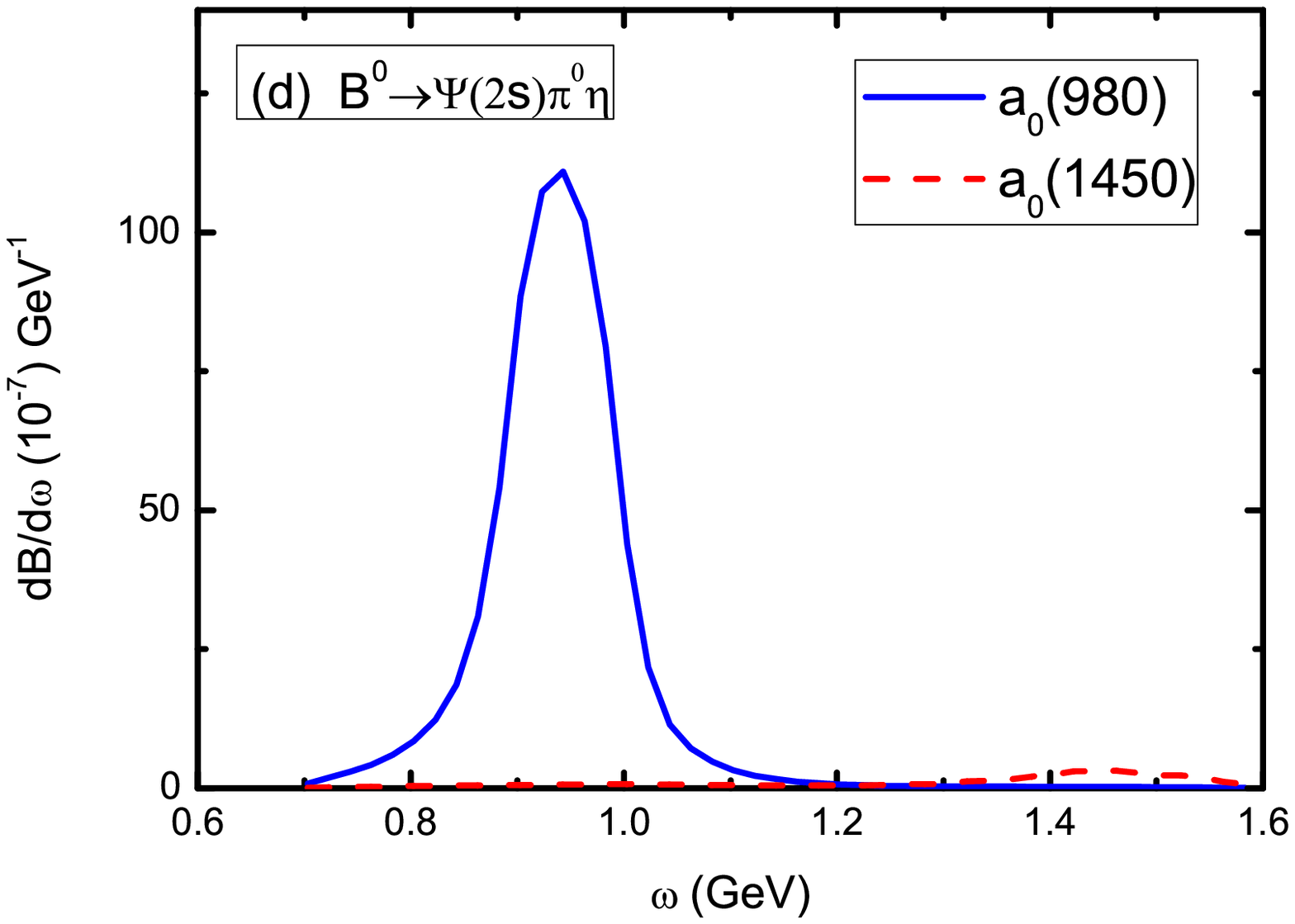}}}
\vspace{-3cm}\caption{
Isovector scalar resonance contributions to the differential branching fractions of
the modes (a) $B^0\rightarrow J/\psi K^+K^-$, (b) $B^0\rightarrow J/\psi \pi^0\eta$,
 (c) $B^0\rightarrow \psi(2S) K^+K^-$, and  (d) $B^0\rightarrow \psi(2S) \pi^0\eta$.
 The  blue solid lines corresponds to the resonant $a_0(980)$ contributions, while the  red dashed to the $a_0(1450)$. }
 \label{fig:pwave1}
\end{center}
\end{figure}

The differential decay branching ratios versus the invariant mass $\omega$ are plotted in Fig. \ref{fig:pwave1}.
Note that the $J/\psi-\psi(2S)$ mass difference causes significant differences in the
range spanned in the respective decay modes.
The blue solid and red dashed curves represent the contributions from the resonances $a_0(980)$ and $a_0(1450)$,
respectively. The different shapes between the two  resonances are mainly governed by the corresponding
 partial amplitude $M_{a_0}$ and  complex parameters $C_{a_0}$ in Eq. (\ref{eq:reso}).
 One can see in Fig. \ref{fig:pwave1} (a) and (c) that a clear narrow peak near the $K^+K^-$ threshold for the $a_0(980)$ resonance,
which makes its distribution to be   suppressed by the phase-space as mentioned before.
The  $a_0(1450)$ resonance peak  has smaller strength than the $a_0(980)$ one,
but its broader width compensate the integrated strength over the full $K^+K^-$ invariant mass.
 Therefore,
 the contributions from the two resonances  are of comparable size for the $K\bar K$ modes [see Table \ref{tab:br}].
 In fact, the Crystal Barrel experiment \cite{prd573860} found the  $a_0(1450)$ component is larger than that
 of $a_0(980)$ resonance in the process of $p\bar p$ annihilation into the $K\bar K\pi$   final state.
 In contrast to the $K\bar K$ channels, the $\pi^0\eta$ threshold   far below   the two resonance poles,
 and the strength of the $\pi^0\eta$ distribution is typical larger than the one for $K\bar K$,
which enhanced  its branching ratio accordingly.
Just as expected, without the additional suppression from the phase space we observe an appreciable strength for $a_0(980)$ excitation and
a less strong, but clearly visible excitation for the $a_0(1450)$ in Fig. \ref{fig:pwave1} (b) and (d).
The obtained distribution for the  $a_0(980)$ resonance contribution to the $B^0\rightarrow J/\psi K^+K^-$ decay agrees fairly
well with the LHCb data shown in Fig. 15 of Ref. \cite{prd88072005},
%while the corresponding $\pi\eta$ channels, the  $a_0(1450)$ resonance contributions,
%while that decays to the $\pi\eta$ pair, the $a_0(1450)$ resonance contributions,
%as well as the  $\psi(2S)$ involved modes
 while other predictions could be tested by future experimental measurements.

\section{ conclusion}\label{sec:sum}

In this work we discuss the isovector scalar resonance contributions to the
three-body $B^0\rightarrow \psi (K\bar K, \pi\eta)$ decays under the quasi-two-body
approximation based on the PQCD framework by introducing the corresponding dimeson DAs.
The involved timelike form factors   are parametrized as  a linear combination of two  components $a_0(980)$ and $a_0(1450)$,
which can be described by the Flatt\'{e} line shape and   Breit-Wigner form, respectively.
The predicted $K^+K^-$ invariant mass distribution as well as its integrated  branching ratio for the $a_0(980)$
resonance in the $B^0\rightarrow J/\psi K^+K^-$ decay
 are in agreement with the findings by the LHCb Collaboration.
It is found that the $a_0(1450)$ contribution is comparable with the $a_0(980)$ one for the $K\bar K$ modes,
while fall short by a large factor for the  $\pi\eta$ sector.
In both  resonances, the strength of the $\pi\eta$ invariant mass distribution are
typical  larger than the $K\bar K$ one in the channels  with the same bachelor charmonia in the final state.
The obtained branching ratios of the $B^0\rightarrow \psi a_0(980)(\rightarrow \pi\eta)$ decays can reach the order of $10^{-6}$,
which would be straightforward for experimental observations.

 We  estimate  the isospin breaking effect, which originates from  the different thresholds of charged and neutral kaons,
  between the two physics final states $K^+K^-$ and $K^0\bar{K}^0$ in the $a_0(980)$ and $a_0(1450)$ channels.
  For the former, the isospin breaking effect can reach roughly $10\%$ even without the $a_0-f_0$  mixing,
  while  for the latter, the isospin breaking effect are negligible since its resonance mass
is far away from the two-kaon thresholds.

We have discussed theoretical uncertainties arising from the
nonperturbative parameters in the initial and final states DAs, and  hard scale.
The nonperturbative parameters   contribute the main uncertainties in our approach,
while the hard scale dependent uncertainty is less than $20\%$ due to the inclusion
of the  vertex corrections. In addition,
the $a_0(980)$ resonance contributions  are largely dependence on the Flatt\'{e} parameters,
which should be constrained in the future.

\begin{acknowledgments}
We thank  Hsiang-nan Li, Wei Wang, and Wen-Fei Wang for useful discussions.
%I would like to acknowledge  for helpful discussions.
This work is supported in part by the National Natural Science Foundation of China
under Grants No.11605060,  and  No.11547020, in part
by the Program for the Top Young Innovative Talents of
Higher Learning Institutions of Hebei Educational
Committee under Grant No. BJ2016041, and in part by Training Foundation
of North China University of Science and Technology under Grant
No. GP201520 and No. JP201512.
\end{acknowledgments}

\end{document}